\documentclass[twocolumn, tighten]{aastex6}

\usepackage{apjfonts}
\usepackage{color}
\usepackage{CJK}

\newcommand{\pivec}{\mbox{\boldmath $\pi$}}

\newcommand{\thetae}{\theta_{\rm E}}
\newcommand{\pie}{\pi_{\rm E}}

\definecolor{darkbrown}{RGB}{139,69,19}

\shorttitle{OGLE-2014-BLG-0289}
\shortauthors{Han et al.}

\begin{document}

\title{OGLE-2014-BLG-0289: Precise Characterization of a Quintuple-Peak Gravitational Microlensing Event}

\author{
A.~Udalski\altaffilmark{01,100}, 
C.~Han\altaffilmark{02,500}, 
V.~Bozza\altaffilmark{03,04,300}, 
A.~Gould\altaffilmark{05,06,400}, 
I.~A.~Bond\altaffilmark{07,200}\\
and\\
P.~Mr\'oz\altaffilmark{01}, 
J.~Skowron\altaffilmark{01}, 
{\L}.~Wyrzykowski\altaffilmark{01}
M.~K.~Szyma\'nski\altaffilmark{01}, 
I.~Soszy\'nski\altaffilmark{01}, 
K.~Ulaczyk\altaffilmark{01}, 
R.~Poleski\altaffilmark{01,06},
P.~Pietrukowicz\altaffilmark{01}, 
S.~Koz{\l}owski\altaffilmark{01}, \\
(The OGLE Collaboration) \\   
F.~Abe\altaffilmark{08}, 
R.~Barry\altaffilmark{09},
D.~P.~Bennett\altaffilmark{09,10}, 
A.~Bhattacharya\altaffilmark{09},
M.~Donachie\altaffilmark{11},
P.~Evans\altaffilmark{11},
A.~Fukui\altaffilmark{12}, 
Y.~Hirao\altaffilmark{13},
Y.~Itow\altaffilmark{08}, 
K.~Kawasaki\altaffilmark{13},
N.~Koshimoto\altaffilmark{13},
M.~C.~A.~Li\altaffilmark{11}, 
C.~H.~Ling\altaffilmark{07}, 
K.~Masuda\altaffilmark{08}, 
Y.~Matsubara\altaffilmark{08}, 
S.~Miyazaki\altaffilmark{13},
H.~Munakata\altaffilmark{08},
Y.~Muraki\altaffilmark{08}, 
M.~Nagakane\altaffilmark{13},
K.~Ohnishi\altaffilmark{14}, 
C.~Ranc\altaffilmark{09},
N.~Rattenbury\altaffilmark{11}, 
T.~Saito\altaffilmark{15}, 
A.~Sharan\altaffilmark{11}, 
D.~J.~Sullivan\altaffilmark{16}, 
T.~Sumi\altaffilmark{13},
D.~Suzuki\altaffilmark{17}, 
P.~J.~Tristram\altaffilmark{18},
T.~Yamada\altaffilmark{13},
A.~Yonehara\altaffilmark{19},\\
(The MOA Collaboration)\\
E.~Bachelet\altaffilmark{20}, 
D.~M.~Bramich\altaffilmark{21,22}, 
G.~D\'Ago\altaffilmark{03,23}, 
M.~Dominik\altaffilmark{24}, 
R.~Figuera~Jaimes\altaffilmark{22,24}, 
K.~Horne\altaffilmark{24}, 
M.~Hundertmark\altaffilmark{25,26}, 
N.~Kains\altaffilmark{27}, 
J.~Menzies\altaffilmark{28}, 
R.~Schmidt\altaffilmark{29}, 
C.~Snodgrass\altaffilmark{30,31}, 
I.~A.~Steele\altaffilmark{32}, 
J.~Wambsganss\altaffilmark{29}\\
(RoboNet Collaboration) \\
R.~W.~Pogge\altaffilmark{06}, 
Y.~K.~Jung\altaffilmark{33}, 
I.-G. Shin\altaffilmark{33}, 
J.~C.~Yee\altaffilmark{33},
W.-T.~Kim\altaffilmark{34},\\
(The $\mu$FUN Collaboration)\\
C.~Beichman\altaffilmark{35},
S.~Carey\altaffilmark{36},
S.~Calchi Novati\altaffilmark{37,03},
W.~Zhu\altaffilmark{39,40} \\
(The {\it Spitzer} Team)
}

\email{cheongho@astroph.chungbuk.ac.kr}

\altaffiltext{01}{Warsaw University Observatory, Al. Ujazdowskie 4, 00-478 Warszawa, Poland} 
\altaffiltext{02}{Department of Physics, Chungbuk National University, Cheongju 28644, Korea} 
\altaffiltext{03}{Dipartimento di Fisica "E. R. Caianiello", Universit\'a di Salerno, Via Giovanni Paolo II, I-84084 Fisciano (SA), Italy} 
\altaffiltext{04}{Istituto Nazionale di Fisica Nucleare, Sezione di Napoli, Via Cintia, I-80126 Napoli, Italy} 
\altaffiltext{05}{Korea Astronomy and Space Science Institute, Daejon 34055, Korea} 
\altaffiltext{06}{Department of Astronomy, Ohio State University, 140 W. 18th Ave., Columbus, OH 43210, USA} 
\altaffiltext{07}{Institute of Natural and Mathematical Sciences, Massey University, Auckland 0745, New Zealand} 
\altaffiltext{08}{Institute for Space-Earth Environmental Research, Nagoya University, 464-8601 Nagoya, Japan} 
\altaffiltext{09}{Code 667, NASA Goddard Space Flight Center, Greenbelt, MD 20771, USA} 
\altaffiltext{10}{Deptartment of Physics, University of Notre Dame, 225 Nieuwland Science Hall, Notre Dame, IN 46556, USA} 
\altaffiltext{11}{Dept.\ of Physics, University of Auckland , Private Bag 92019, Auckland, New Zealand} 
\altaffiltext{12}{Okayama Astrophysical Observatory, National Astronomical Observatory of Japan, Asakuchi,719-0232 Okayama, Japan} 
\altaffiltext{13}{Department of Earth and Space Science, Graduate School of Science, Osaka University, Toyonaka, Osaka 560-0043, Japan}
\altaffiltext{14}{Nagano National College of Technology, 381-8550 Nagano, Japan} 
\altaffiltext{15}{Tokyo Metroplitan College of Industrial Technology, 116-8523 Tokyo, Japan} 
\altaffiltext{16}{School of Chemical and Physical Sciences, Victoria University, Wellington, New Zealand} 
\altaffiltext{17}{Institute of Space and Astronautical Science, Japan Aerospace Exploration Agency, Kanagawa 252-5210, Japan}
\altaffiltext{18}{Mt. John University Observatory, P.O. Box 56, Lake Tekapo 8770, New Zealand} 
\altaffiltext{19}{Department of Physics, Faculty of Science, Kyoto Sangyo University, 603-8555 Kyoto, Japan}
\altaffiltext{20}{Las Cumbres Observatory Global Telescope Network, 6740 Cortona Drive, Suite 102, Goleta, CA 93117, USA} 
\altaffiltext{21}{New York University Abu Dhabi, Saadiyat Island, Abu Dhabi, PO Box 129188, United Arab Emirates} 
\altaffiltext{22}{European Southern Observatory, Karl-Schwarzschild-Str. 2, 85748 Garching bei M unchen, Germany} 
\altaffiltext{23}{Istituto Nazionale di Fisica Nucleare, Sezione di Napoli, Napoli, Italy} 
\altaffiltext{24}{School of Physics \& Astronomy, University of St Andrews, North Haugh, St Andrews KY 16 9SS, UK} 
\altaffiltext{25}{Niels Bohr Institute, University of Copenhagen, Juliane Maries Vej 30, 2100, Kobenhavn, Denmark} 
\altaffiltext{26}{SUPA, School of Physics \& Astronomy, University of St Andrews, North Haugh, St Andrews KY16 9SS, UK} 
\altaffiltext{27}{Space Telescope Institute, 3700 San Martin Drive, Baltimore, MD 21218, USA} 
\altaffiltext{28}{South African Astronomical Observatory, PO Box 9, Observatory 7935, South Africa} 
\altaffiltext{29}{Astronomisches Rechen-Institut, Zentrum f\"ur Astronomie der Universit at Heidelberg (ZAH), 69120 Heidelberg, Germany} 
\altaffiltext{30}{Planetary and Space Sciences, Dept of Physical Sciences, The Open University, Milton Keynes, MK7 6AA, UK} 
\altaffiltext{31}{Max Planck Institute for Solar System Research, Justus-von-Liebig-Weg 3, 37077 Gotingen, Germany} 
\altaffiltext{32}{Astrophysics Research Institute Liverpool John Moores University, Liverpool L3 5RF, UK} 
\altaffiltext{33}{Harvard-Smithsonian Center for Astrophysics, 60 Garden St., Cambridge, MA, 02138, US} 
\altaffiltext{34}{Department of Physics \& Astronomy, Seoul National University, Seoul 151-742, Korea} 
\altaffiltext{35}{NASA Exoplanet Science Institute, California Institute of Technology, Pasadena, CA 91125, USA}
\altaffiltext{36}{Spitzer Science Center, MS 220-6, California Institute of Technology, Pasadena, CA, USA}
\altaffiltext{37}{IPAC, Mail Code 100-22, California Institute of Technology, 1200 E. California Boulevard, Pasadena, CA 91125, USA}
\altaffiltext{39}{Department of Astronomy, Ohio State University, 140 W. 18th Ave., Columbus, OH 43210, USA}
\altaffiltext{40}{Canadian Institute for Theoretical Astrophysics, 60 St George Street, University of Toronto, Toronto, ON M5S 3H8, Canada}
\altaffiltext{100}{OGLE Collaboration}
\altaffiltext{200}{The MOA Collaboration}
\altaffiltext{300}{The RoboNet collaboration}
\altaffiltext{400}{The $\mu$FUN Collaboration}
\altaffiltext{500}{Corresponding author}

\begin{abstract}
We present the analysis of the binary-microlensing event OGLE-2014-BLG-0289. The 
event light curve exhibits very unusual five peaks where four peaks were 
produced by caustic crossings and the other peak was produced by a cusp approach.  
It is found that the quintuple-peak features of the light curve provide tight 
constraints on the source trajectory, enabling us to precisely and accurately 
measure the microlensing parallax $\pi_{\rm E}$.  Furthermore, the three resolved 
caustics allow us to measure the angular Einstein radius $\thetae$.  From the 
combination of $\pi_{\rm E}$ and $\thetae$, the physical lens parameters are 
uniquely determined. It is found that the lens is a binary composed of two M dwarfs 
with masses $M_1 = 0.52 \pm 0.04\ M_\odot$ and $M_2=0.42 \pm 0.03\ M_\odot$ separated 
in projection by $a_\perp = 6.4 \pm 0.5$ au. The lens is located in the disk with a 
distance of $D_{\rm L} = 3.3 \pm 0.3$~kpc.  It turns out that the reason for the 
absence of a lensing signal in the {\it Spitzer} data is  that the time of observation 
corresponds to the flat region of the light curve. 
\end{abstract}

\keywords{gravitational lensing: micro -- binaries: general}

\section{Introduction}
 
Since commencing in the early 1990s \citep{Udalski1994, Alcock1995, Aubourg1995}, massive 
surveys have detected numerous microlensing events. The detection rate of microlensing 
events, which was of order $10\ {\rm yr}^{-1}$ in the early stage of the surveys, has 
greatly increased and currently more than 2000 events are annually detected.

However, determinations of lens masses have been possible for very limited cases.
The difficulty of the lens mass measurement arises because the event timescale, 
which is the only measurable quantity related to the lens mass for general lensing 
events, is related to not only the mass $M$ but also to the relative lens-source 
proper motion $\mu$ and the lens-source parallax $\pi_{\rm rel}$, i.e. 
\begin{equation}
t_{\rm E}={\sqrt{\kappa M \pi_{\rm rel}}\over \mu};\qquad
\pi_{\rm rel}={\rm au}\left( {1\over D_{\rm L}}-{1\over D_{\rm S}}  \right),
\end{equation}
where $\kappa=4G/(c^2{\rm au})$ and $D_{\rm L}$ and $D_{\rm S}$ represent the 
distances to the lens and source, respectively.  For the unique determination of 
the lens mass, one needs to measure two additional observables of the angular 
Einstein radius $\thetae$ and the microlens parallax $\pie$, i.e.
\begin{equation}
M={\thetae \over \kappa\pie },
\end{equation}
where $\thetae^2=\kappa M \pi_{\rm rel}$ \citep{Gould2000}.

The angular Einstein radius can be measured by detecting light curve deviations  
caused by finite-source effects \citep{Gould1994, Nemiroff1994}.  For lensing 
events produced by single masses, finite-source effects can be detected when a 
lens crosses the surface of a source star \citep{Pratt1996, Choi2012}. However, 
the ratio of the angular source radius $\theta_*$ to the angular Einstein radius 
is of order $10^{-3}$ for a main-sequence source star and of order  $10^{-2}$ even 
for a giant star.  Therefore, the chance to detect finite-source effects for a 
single-lens event is very low.  For events produced by binary objects, on the other 
hand, the probability of $\thetae$ measurement is relatively high because binary-lens 
events usually produce caustic-crossing features from which finite-source effects 
can be detected.

One can measure the microlens parallax from the light curve deviation induced by 
the acceleration of the source motion caused by the Earth's orbital motion: 
`annual microlens parallax' \citep{Gould1992}.  One can also measure the microlens 
parallax by simultaneously observing a lensing event from ground and from a satellite 
in a heliocentric orbit: `space-based microlens parallax' \citep{Refsdal1966, 
Gould1994}.  Considering that the physical Einstein radius 
$r_{\rm E} = D_{\rm L}\thetae$ of typical Galactic lensing events is of order a few au, 
for satellites with a projected Earth-satellite separation of order au, the light 
curves seen from the Earth and the satellite usually exhibit considerable differences, 
e.g.\ 
OGLE-2015-BLG-0124 \citep{Udalski2015},
OGLE-2015-BLG-0966 \citep{Street2016}, OGLE-2015-BLG-1268, and OGLE-2015-BLG-0763 
\citep{Zhu2016}, and this enables a precise measurement of $\pie$.  In contrast, 
deviations in lensing light curves induced by annual microlens-parallax effects are 
in most cases very subtle due to the small positional change of the Earth during 
$\sim (O)10$ day durations of typical lensing events.  Furthermore, parallax-induced 
deviations can often be confused with deviations caused by other higher-order effects 
such as the orbital motion of the lens \citep{Batista2011, Skowron2011, Han2016b}.  
As a result, measurements of  annual microlens parallaxes are in many cases subject 
to large uncertainty both in precision and accuracy.

It was pointed out by \citet{An2001} that the chance to determine the lens mass by 
measuring both $\pie$ and $\thetae$ is high for a subclass of binary lensing events 
with three well-measured peaks where two peaks are produced by caustic crossings and 
the other is produced by a cusp approach. This is because the individual peaks provide 
tight constraints on the source trajectory, enabling one to measure the microlens 
parallax. Furthermore, the angular Einstein radius is measurable from the analysis 
of almost any well-resolved caustic crossing, making triple-peak events good candidates 
for lens mass measurements.

In this paper, we present the analysis of the binary-lens event OGLE-2014-BLG-0289. 
The light curve of the event exhibits very unusual five peaks. Among these peaks, 
four were produced by caustic crossings and the other was produced by a cusp approach. 
The angular Einstein radius is precisely measured by detecting finite-source effects 
from the resolutions of 3 caustic crossings. Furthermore, the well-resolved multiple 
peaks enable us to measure the microlens parallax, leading to an accurate and precise 
measurement of the lens mass.

\section{Observation and Data}
 
\begin{figure}
\includegraphics[width=\columnwidth]{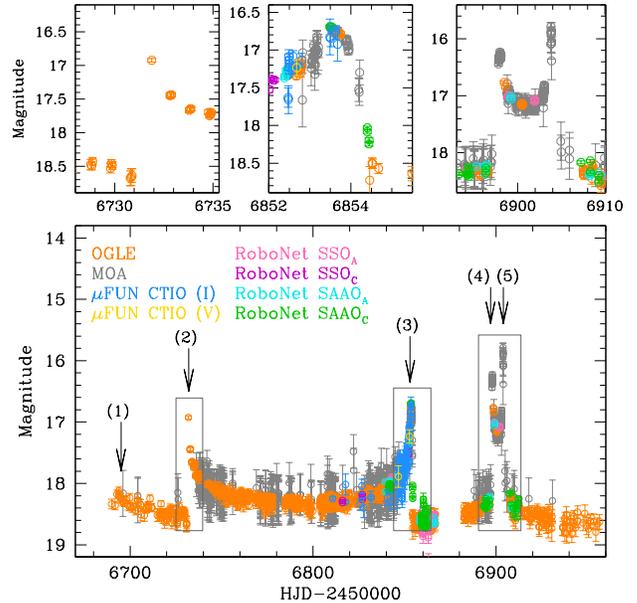}
\caption{
Light curve of OGLE-2014-BLG-0289. The upper panels show the zoom of the regions 
enclosed by boxes in the lower panel. The numbered arrows designate the peaks 
in the light curve. The numbered arrows indicate the locations of the five peaks.}
\label{fig:one}
\end{figure}

The source star of the microlensing event OGLE-2014-BLG-0289 is located toward the 
Galactic bulge field.  The equatorial coordinates of the source star are 
$({\rm RA},{\rm DEC})_{\rm J2000}=$ (17:53:51.66, -29:05:05.6), which 
correspond to Galactic coordinates 
$(l,b)=(0.80^\circ, -1.62^\circ)$. 
The magnification of the source flux caused by lensing was found on 17 March 2014 
(${\rm HJD}'={\rm HJD}-2450000\sim 6733$) by the Early Warning System of the Optical 
Gravitational Lensing Experiment \citep[OGLE:][]{Udalski2003} survey from observations 
conducted using the 1.3 m Warsaw Telescope at the Las Campanas Observatory in Chile. 
The event was also in the footprint of the Microlensing Observations in Astrophysics 
\citep[MOA:][]{Bond2001, Sumi2003} survey that was conducted using the 1.8 m telescope 
at Mt.\ John Observatory in New Zealand.  The event was dubbed MOA-2014-BLG-092 in the 
``MOA Transient Alerts'' list 
\footnote{http://www.massey.ac.nz/~iabond/moa/alerts/listevents.php?year=2014}. 
Data from the OGLE and MOA surveys were acquired in the the standard Cousins $I$
and the customized MOA $R$ passband, respectively.
Figure~\ref{fig:one} shows the light curve of the event.

The source flux was already magnified before the 2014 Bulge season started.  
Just one day before the event was identified, i.e.\ ${\rm HJD}'\sim 6732$, the 
light curve exhibited a sharp spike.  Such a spike feature commonly appears in binary 
lensing events and is produced when a source crosses the caustic of a binary lens.  
Caustics indicate source positions at which the flux of a point source is infinitely 
magnified.  Caustic crossings in binary-lens events occur in pairs because binary 
caustics form closed curves.  When the source passes the inner region of a caustic, 
the light curve exhibits a characteristic U-shape trough followed by another spike 
that occurs when the source exits the caustic. The event followed this U-shape 
pattern of a binary lensing event until another spike appeared at ${\rm HJD}'\sim 6854$.

The event was analyzed in real time with its progress. On 18 June 2014 
(${\rm HJD}'\sim 6827$), when the event passed the bottom of the U-shape trough, 
the first model was announced to the microlensing community by C.~Han and V.~Bozza.  
According to this model, the spike in the light curve was produced by the crossing 
of the source over the single big caustic formed by a lens that is consisted of two 
similar masses separated in projection by $\sim \thetae$.  Real-time modeling was 
important in preparing follow-up observations to resolve the caustic crossing, 
which yields the angular Einstein radius. It also helped to prepare space-based 
observations using {\it Spitzer} telescope, which was separated $\sim 1$ au from 
the Earth at the time of the event.  Caustic-crossing binary-lens events are 
important targets of {\it Spitzer} observations because one can measure the 
space-based microlens parallax and the measured $\pie$ combined with $\thetae$ 
leads to the measurement of the lens mass.  Due to these considerations, real-time 
modeling was conducted more frequently as the source approached closer to the 
caustic exit.

The caustic exit occurred at ${\rm HJD}'\sim 6854$ which approximately matched the time 
predicted by modeling.  Follow-up observation were conducted by two groups including the 
Microlensing Follow-Up Network ($\mu$FUN) and RoboNet. The $\mu$FUN group observed the event 
using the 1.3 m SMART telescope at the CTIO Observatory in Chile. The RoboNet observations 
were conducted with the 1 m robotic telescopes at South African Astronomical Observatory 
(SAAO) in South Africa and Siding Spring Observatory (SSO) in Australia.  $\mu$FUN 
observations  were conducted in standard Cousins $I$ band and several $V$-band images were 
obtained to measure the source color.  RoboNet data were taken in SDSS-i band.  From these 
follow-up observations, the second spike produced by the source star's caustic exit was 
captured with sufficient resolution to determine the angular source size. We note that 
the caustic exit was also covered by both OGLE and MOA surveys. See the upper middle 
panel of Figure~\ref{fig:one}.

The event continued after the caustic exit and so did real-time modeling. Modeling 
conducted several days after the caustic exit revealed two important findings. First, 
it was predicted that there would be another pair of caustic crossings.  Second, it 
was found that considering the microlens-parallax effect is important for the precise 
description of the observed light curve.  With the progress of the event, the time of 
the next caustic crossing was refined. The predicted time of the caustic crossing was 
informed to the microlensing community and follow-up observations were prepared accordingly.  
The third and fourth caustic crossings occurred successively at ${\rm HJD}'\sim 6897$ and 
6904, respectively.  Although these crossing were missed by follow-up observations, they 
were partially captured by MOA survey observations.  See the upper right panel in 
Figure~\ref{fig:one}.  After these caustic crossings, the event gradually returned 
its baseline magnitude of $I\sim 18.75$.

{\it Spitzer} observations \citep{Calchi2015} were conducted during the period 
$6814 < {\rm HJD}' < 6848$, which corresponded to the time when the event was in the 
trough region between the first two caustic crossings as seen from the ground.  From 
the photometry of the {\it Spitzer} data, however, it is found that there exists no 
noticeable lensing signal, i.e.~no variation of the source brightness.  We discuss 
the reason for the absence of the {\it Spitzer} lensing signal in Section 3.

A very unusual characteristic of the event is that the light curve exhibits 5 peaks. 
The individual peaks occurred at ${\rm HJD}'\sim 6698$, 6732, 6854, 6897, and 6904, 
which are marked by the numbered arrows in Figure~\ref{fig:one}. Based on the shapes 
of the peaks, the first peak appears to be produced by the source's cusp approach, 
while the other 4 peaks were produced by caustic crossings. Multiple sets of caustic 
crossings can occur when a source trajectory asymptotically passes a fold of a 
concave caustic. It turns out that the existence of the multiple peaks in the light 
curve enables precise characterization of the lens system.

We conduct photometry of the data using the pipelines of the individual observation 
groups.  The OGLE  \citep{Wozniak2000, Udalski2003}  and MOA \citep{Bond2001} 
pipelines are based on the Difference Image Analysis (DIA) technique \citep{Alard1998} 
and customized by the individual groups.  The RoboNet and $\mu$FUN data were reduced 
with the DANDIA pipeline \citep{Bramich2008} and the pySIS \citep{Albrow2009}, 
respectively.  For the $\mu$FUN CTIO data, photometry were additionally done with 
DoPHOT software \citep{Schechter1993} in order for the source color measurement and 
color-magnitude diagram construction.  We note that the quality of the MOA data at 
the baseline is not good, but their coverage of the caustic crossings is important 
in measuring $\thetae$.  We, therefore, use MOA data taken when the source was magnified.

For the use of multiple data sets that are obtained with different telescopes 
and detectors and processed with different photometry softwares, it is required 
to readjust the errorbars of the data sets.  For this readjustment, we follow the 
standard procedure of \citet{Yee2012}, where the error bars are renormalized by 
\begin{equation}
\sigma= k (\sigma_0^2+\sigma_{\rm min}^2)^{1/2},
\end{equation}
where $\sigma_0$ is the uncorrected error bar from the automated pipelines. 
We set the factor $\sigma_{\rm min}$ based on the scatter of data.  The factor 
$k$ is set so that $\chi^2$ per degree of freedom (dof) becomes unity, i.e.\ 
$\chi^2/{\rm dof}=1$.  We list the error-bar readjustment factors in 
Table~\ref{table:one} along with the number of data points, $N_{\rm data}$.

\begin{deluxetable}{lrrr}
\tablecaption{Error bar readjustment factors \label{table:one}}
\tablewidth{0pt}
\tabletypesize{\small}
\tablehead{
\multicolumn{1}{c}{Data set} &
\multicolumn{1}{c}{$k$}  &
\multicolumn{1}{c}{$\sigma_{\rm min}$}  &
\multicolumn{1}{c}{$N_{\rm data}$}  
}
\startdata                                              
OGLE                     &  1.808   &    0.002 &  3700  \\
MOA                      &  1.252   &    0.003 &  808   \\
$\mu$FUN CTIO            &  1.122   &    0.005 &  59    \\
RoboNet SSO (Dome A)     &  0.692   &    0.025 &  61    \\
RoboNet SSO (Dome C)     &  0.824   &    0.005 &  32    \\
RoboNet SAAO (Dome A)    &  0.663   &    0.020 &  47    \\
RoboNet SAAO (Dome C)    &  0.583   &    0.020 &  82   
\enddata                                              
\end{deluxetable}

\section{Light Curve Modeling}

From the spike features, it is obvious that the event was produced by a lens composed 
of multiple components. We, therefore, start modeling of the observed light curve 
based on the binary-lens interpretation.  For the simplest case where the relative 
lens-source motion is rectilinear, one needs 7 principal parameters in order to 
describe the light curve of a binary-lens event.  The first three parameters 
$(t_0, u_0, t_{\rm E})$ are needed to describe the source approach to the lens and 
they represent the time of the closest lens-source separation, the separation at 
that time, and the event timescale, respectively.  Another three parameters 
$(s, q, \alpha)$ are used to describe the binary lens and they denote the binary 
separation, mass ratio between the lens components, and the angle between the source 
trajectory and the line connecting the binary components, respectively.  The 
caustic-crossing parts of a binary-lens event are affected by finite-source effects 
and the last parameter $\rho$ is used to describe the deviation.

The light curve of the event exhibits caustic-crossing features and thus we consider 
finite-source effects.  We compute lensing magnifications affected by finite-source 
effects using the inverse ray-shooting technique.  In computing finite-source 
magnifications, we take the surface brightness variation caused by limb-darkening 
into consideration.  The surface-brightness profile is approximated by 
a linear model, i.e. 
\begin{equation}
\Sigma_\lambda \propto 1- \Gamma_\lambda \left( 1- {3\over 2}\cos \phi \right),
\end{equation}
where $\lambda$ denotes the observed passband, $\Gamma_\lambda$ is the 
linear limb-darkening coefficient, and $\phi$ represents the angle between 
the line of sight and the normal to the surface of the source star. 
We determine the limb-darkening coefficients based on the source star's stellar type. 
It turns out that the source is an early K-type main-sequence 
star. See Section 4 for the detailed procedure of the source type determination. 
Based on the stellar type, we adopt the limb-darkening coefficients from the 
\citet{Claret2000} catalog.
The adopted $I$- and $V$-band coefficients are
$\Gamma_I=0.485$ and $\Gamma_V=0.676$, respectively. 
For the MOA $R$-band data, we 
use $\Gamma_{\rm MOA}=(\Gamma_I+\Gamma_R)/2=0.535$, where $\Gamma_R=0.585$ is the 
$R$-band coefficient.

We search for the solution of the lensing parameters in two steps.  In the first 
step, we divide the lensing parameters into two groups.  We select $(s, q, \alpha)$ 
as grid parameters since lensing magnifications can vary dramatically with the small 
change of these parameters.  We choose the other parameters, i.e.\ 
$(t_0, u_0, t_{\rm E}, \rho)$, as downhill parameters because lensing magnifications 
vary smoothly with the changes of the parameters.  For the individual sets of the 
grid parameters, we then search for the set of the downhill parameters yielding the 
best $\chi^2$ using the Markov Chain Monte Carlo (MCMC) method.  The total computation 
time for the grid search is $\sim 24$ hours using 176 CPUs.  This initial search 
provides a $\chi^2$ map in the $s$-$q$-$\alpha$ 
parameter space, from which we identify local minima.  We then refine each local minimum 
by allowing all parameters to vary. We note that the initial grid search is important to 
identify degenerate solutions where different combinations of lensing parameters result 
in similar lensing light curves. For the case of OGLE-2014-BLG-0289, we identify a unique 
solution and find no solution with $\chi^2$ comparable to the best-fit solution.

\begin{figure}
\includegraphics[width=\columnwidth]{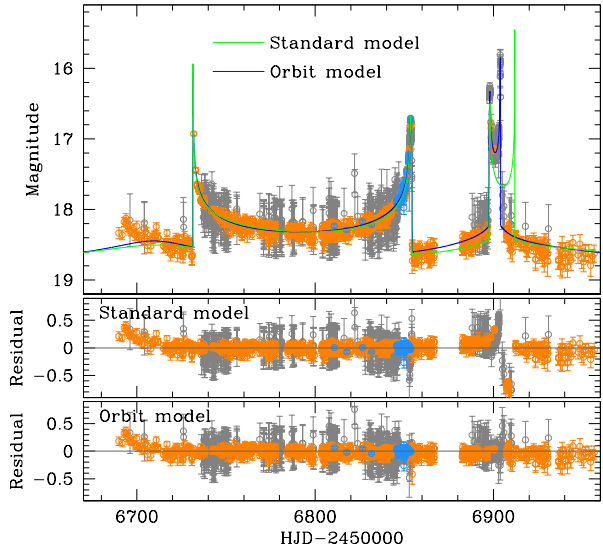}
\caption{
Model light curves of the ``standard'' (blue curve) and ``orbit'' (red curve) 
solutions. The middle and lower panels show the residuals from the individual 
models. 
}
\label{fig:two}
\end{figure}

In Figure~\ref{fig:two}, we present the model light curve (blue curve in the 
upper panel) of the solution obtained under the assumption of the rectilinear 
lens-source motion (``standard model'').  The middle panel shows the residual 
from the model.  For better visual comparison of the fit with data, we plot data 
points of only the OGLE, MOA, and $\mu$FUN CTIO data sets.  The binary-lens 
parameters estimated by the model are $s\sim 1.6$ and $q\sim 0.9$.  Although the 
standard model basically describes the overall light curve, it leaves considerable 
residuals. The major residuals occur near the first peak at ${\rm HJD}'\sim 6700$ 
and the fifth peak at ${\rm HJD}'\sim 6904$.

The inconsistency of the model with the data suggests the need to consider 
higher-order effects.  Considering the long duration of the event, 
it is suspected that the assumption of a rectilinear lens-source 
motion may not be valid. There exist two major effects that can cause deviations 
of the relative lens-source motion from rectilinear. One such effect is caused by 
the orbital motion of the Earth around the sun, i.e.\ microlens-parallax effect.  
The other is caused by the orbital motion of the binary lens itself, lens-orbital 
effect. We, therefore, check whether the residuals from the standard model
can be explained by these higher-order effects.

\begin{deluxetable}{lc}
\tablecaption{Comparison of Models \label{table:two}}
\tablewidth{0pt}
\tabletypesize{\small}
\tablehead{
\multicolumn{1}{c}{Model} &
\multicolumn{1}{c}{$\chi^2$}  
}
\startdata                                              
Standard                  & 17950.3    \\
Orbit                     & 5011.8    \\
Parallax ($u_0>0$)        & 4847.2     \\
Parallax ($u_0<0$)        & 4867.0     \\
Parallax+Orbit ($u_0>0$)  & 4829.9     \\    
Parallax+Orbit ($u_0<0$)  & 4848.0 
\enddata                                              
\end{deluxetable}

Incorporating higher-order effects requires to include additional lensing parameters. 
In order to consider the microlens-parallax effect, one needs 2 parameters 
$\pi_{{\rm E},N}$ and $\pi_{{\rm E},E}$.  They denote the north and east components 
$\pivec_{\rm E}$ that represents the microlens-parallax vector projected onto the 
sky in the equatorial coordinate systems.  The direction of $\pivec_{\rm E}$ is the 
same as that of the relative lens-source motion \citep{Gould2000, Gould2004}.  Under 
the first-order approximation that the projected binary separation $s$ and the source 
trajectory angle $\alpha$ vary in constant rates, the lens-orbital effect is described 
by 2 parameters of $ds/dt$ and $d\alpha/dt$ \citep{Albrow2000}.  With 
these parameters, we conduct additional modeling to check the improvement of the 
fit with the higher-order effects. In this modeling, we first separately consider 
the microlens-parallax (``parallax model'') and lens-orbital effects (``orbit model'') 
and second simultaneously consider both effects (``parallax + orbit'' model).

For events affected by microlens-parallax effects, there may exist a pair of 
degenerate solutions with $u_0>0$ and $u_0<0$: `ecliptic degeneracy' \citep{Skowron2011}. 
This degeneracy arises because the source trajectories of the two degenerate solutions
are in the mirror symmetry with respect to the binary-lens axis.
For the pair of the solutions subject to this degeneracy, the lensing parameters 
are approximately related by
$(u_0, \alpha, \pi_{{\rm E},N}, d\alpha/dt) \leftrightarrow 
-(u_0, \alpha, \pi_{{\rm E},N}, d\alpha/dt)$. We check this degeneracy whenever 
microlens-parallax effects are considered in modeling.

\begin{figure}
\includegraphics[width=\columnwidth]{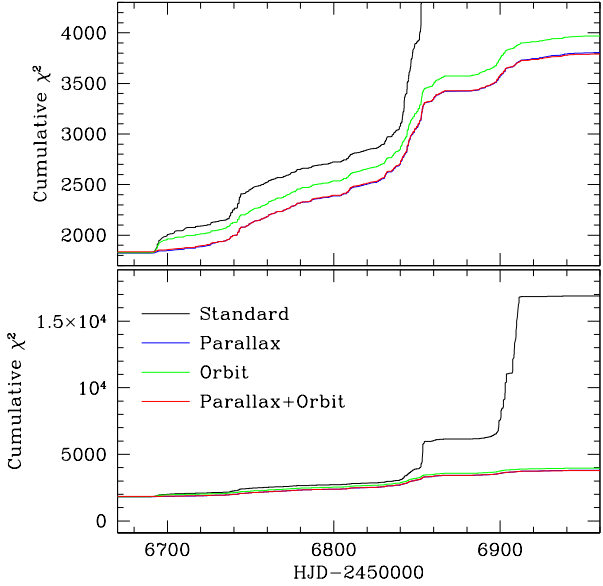}
\caption{
The cumulative distributions of $\chi^2$ as a function of time for the tested models.  
To better show the differences between models considering higher-order effects,
we present the  zoom of the distributions in the upper panel.
We note that the $\chi^2$ difference between the ``parallax'' and ``parallax+orbit'' 
models is so small that the two distributions are difficult to be distinguished within 
the line width.
}
\label{fig:three}
\end{figure}

We find that higher-order effects, particularly the microlens-parallax effect, 
are important in explaining the residuals from the standard model.  In Table~\ref{table:two}, 
we present the $\chi^2$ values of the tested models.  In Figure~\ref{fig:three}, we also 
present the cumulative distributions of $\chi^2$ as a function of time for the individual 
models.  From the comparison of models, it is found that the fit improves by $\Delta\chi^2 
\sim 13103.1$ and 12938.5 by the microlens-parallax and lens-orbital effects, respectively, 
with respect to  the standard model.  Among the two higher-order effects, it is found that 
the microlens-parallax effect is the main cause of the deviation from the standard model.  
The dominance of the microlens-parallax effect over the lens-orbital effect is found from 
the facts that (1) the ``parallax'' model yields substantially better fit than the ``orbit'' 
model (by $\Delta\chi^2=164.6$), (2) the parallax-only model can describe all five peaks, 
and  (3) the further improvement from the parallax fit with the additional consideration 
of the lens-orbital effect ($\Delta\chi^2=17.3$) is minor.

In Figure~\ref{fig:two}, we present the model light curve the orbit model (red 
curve in the upper panel) and the residual from the model (lower panel).  It is found 
the model can describe the fifth peak, that could not be explained by the standard model, 
but it still cannot describe the first peak.  To check the possibility that the assumption 
of the constant change rates of $ds/dt$ and $d\alpha/dt$ do not sufficiently describe 
lens-orbital effects, we conduct an additional modeling by fully considering the Keplerian 
orbital motion of the lens. This modeling requires 2 more parameters of $s_\parallel$ 
and $ds_\parallel/dt$.  These parameters represent binary separation (in units of $\thetae$) 
along the line of sight and the rate of separation change, respectively \citep{Skowron2011, Shin2011}.  
This modeling results in almost an identical $\chi^2$ to that of the linear 
orbital-motion solution.  This confirms that the major cause of the deviation is the 
microlens-parallax effect.

\begin{figure}
\includegraphics[width=\columnwidth]{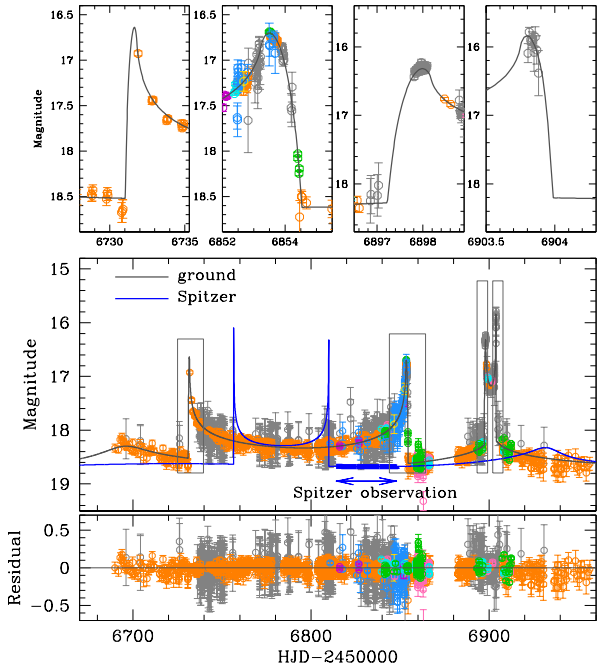}
\caption{
Best-fit model light curve (black solid curve). Upper panels show the model fits around 
the regions enclosed by boxes in the middle panel.  The lower panel shows the residual 
from the model.  The blue curve in the middle panel represents the light curve expected 
to be observed in space using {\it Spitzer} telescope.  The region represented by a 
left-right arrow and marked by `{\it Spitzer} observation'  denotes the period during 
which {\it Spitzer} observations were conducted ($2456814 < {\rm HJD} < 2456848$). 
}
\label{fig:four}
\end{figure}

In Table~\ref{table:three}, we list the lensing parameters of the $u_0>0$ and $u_0<0$ 
solutions of the ``parallax+orbit'' model. Also presented are the fluxes of the source, 
$F_s$, and the blended light, $F_b$, that are measured based on the OGLE data.  From 
the comparison of $u_0>0$ and $u_0<0$ solutions, it is found that the $u_0>0$ solution 
is slightly preferred over $u_0<0$ solution by $\Delta\chi^2=18.1$.  In Figure~\ref{fig:four}, 
we present the model light curve of the best-fit solution (``parallax+orbit'' with $u_0>0$).  
To better show the fits around the caustic-crossing features, we also present the zoom of 
the regions in the upper panels. It is found that the model precisely describes all peaks.

\begin{deluxetable}{lcc}
\tablecaption{Best-fit Lensing Parameters \label{table:three}}
\tablewidth{0pt}
\tabletypesize{\small}
\tablehead{
\multicolumn{1}{c}{Parameter} &
\multicolumn{2}{c}{Value}  \\
\multicolumn{1}{l}{} &
\multicolumn{1}{c}{$u_0>0$} &
\multicolumn{1}{c}{$u_0<0$}  
}
\startdata                                              
$\chi^2$                 & 4829.9                 & 4848.8                \\
$t_0$ (HJD')             & $6820.781 \pm 0.327 $  & $6802.729 \pm 0.391$  \\
$u_0$                    & $0.049    \pm 0.001 $  & $-0.031   \pm 0.001$  \\
$t_{\rm E}$ (days)       & $144.43   \pm 0.24  $  & $162.59   \pm 1.40 $  \\ 
$s$                      & $1.64     \pm 0.01  $  & $1.59     \pm 0.01 $  \\
$q$                      & $0.81     \pm 0.01  $  & $1.06     \pm 0.01 $  \\
$\alpha$ (rad)           & $2.852    \pm 0.004 $  & $-2.739   \pm 0.004$  \\
$\rho$ ($10^{-3}$)       & $0.52     \pm 0.01  $  & $0.57     \pm 0.01 $  \\
$\pi_{{\rm E},N}$        & $0.111    \pm 0.002 $  & $-0.071   \pm 0.005$  \\
$\pi_{{\rm E},E}$        & $-0.104   \pm 0.002 $  & $-0.051   \pm 0.006$  \\
$ds/dt$ (yr$^{-1}$)      & $-0.04    \pm 0.02  $  & $-0.01    \pm 0.01 $  \\
$d\alpha/dt$(yr$^{-1}$)  & $0.10     \pm 0.01  $  & $-0.07    \pm 0.01 $  \\  
$(F_s/F_b)_{\rm OGLE}$   & 0.097/0.392            & 0.108/0.380         
\enddata                                              
\end{deluxetable}

\begin{figure}
\includegraphics[width=\columnwidth]{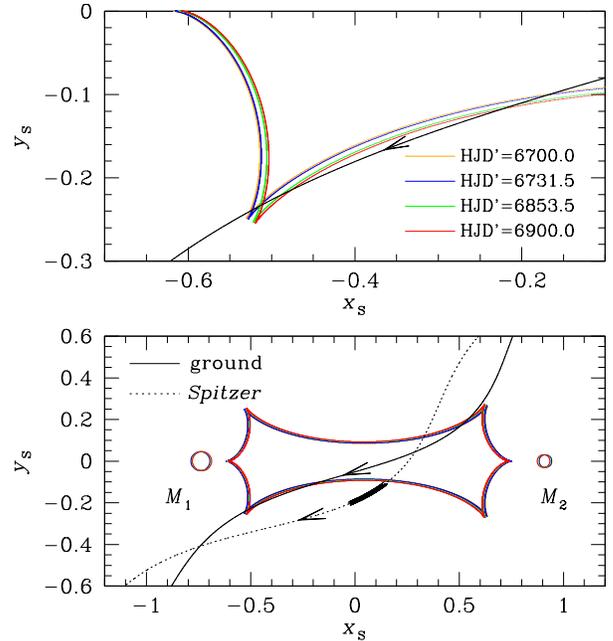}
\caption{
Lens-system geometry showing the source trajectory (solid curve with an arrow) 
with respect to the caustic (cuspy closed curve) and the lens components (marked
by $M_1$ and $M_2$).  The upper panel shows the enlargement of the lower left region 
of the caustic.  The caustics at 4 different times are presented in different colors.  
The dotted curve represents the source trajectory seen in space from the {\it Spitzer} 
telescope.  The thick line on the {\it Spitzer} source trajectory represents the 
time during which the event was observed by the {\it Spitzer} telescope.
}
\label{fig:five}
\end{figure}

Figure~\ref{fig:five} shows the lens-system geometry.  In the geometry, we present 
the source trajectory (solid curve with an arrow) with respect to the caustic (cuspy 
closed curve) and the lens components (marked by $M_1$ and $M_2$) for the best-fit 
solution, i.e.\ $u_0>0$ solution of the parallax+orbit model.  To show the variation 
of the lens positions and the resulting caustic due to lens-orbital effects, we present 
the caustics 
corresponding to 4 different times of the caustic crossings, although it is difficult 
to see the variation due to the minor lens-orbital effects.  From the geometry, it is 
found that the first peak (at ${\rm HJD}'\sim 6698$), which could be explained neither 
by the standard model nor by the orbital model, is explained by the source star's approach 
close to the cusp on the upper right part of the caustic. Being curved by the 
microlens-parallax effect, the source trajectory passes an outer edge of the lower left 
cusp and the model can describe the last peak, which was not be able to be explained by 
the standard model.

We find that the absence of lensing signals in the {\it Spitzer} data is due to the 
fact that the space-based light curve during the {\it Spitzer} observation accidentally 
corresponds to a region where the light curve is very flat.  In the middle panel of 
Figure~\ref{fig:four}, we present the light curve expected to be observed in space using 
the {\it Spitzer} telescope  (blue curve).  We note that the {\it Spitzer} light curve 
is constructed based on the microlens parallax parameters determined from the ground-based 
data.  The region represented by a left-right arrow and marked by `{\it Spitzer} observation' 
denotes the period during which {\it Spitzer} observations were conducted.  It shows that 
this region of the light curve is very flat and thus there is no noticeable lensing signal 
in the {\it Spitzer} data.  In the lower panel of Figure~\ref{fig:five}, we present the 
source trajectory (dotted curve with an arrow) that is expected to be seen from the 
{\it Spitzer} telescope.  The consistency of the predicted model with the flat {\it Spitzer} 
data further supports the correctness of the solution determined from the ground-based data.

If the source is a binary, the orbital motion of the source can also induce long-term 
deviations in lensing light curves: ``xallarap effect'' \citep{Poindexter2005, Rahvar2009}.  
We, therefore, check the xallarap possibility of the deviation.  Considering xallarap effects 
requires five parameters in addition to the principal parameters.  These include the north and 
east components of the xallarap vector, $\xi_{{\rm E},N}$ and $\xi_{{\rm E},E}$, the orbital 
period, the phase angle and inclination of the orbit.  See the appendix of \citet{Han2016a} 
for details about the xallarap parameters.  We find a best-fit xallarap model with an orbital 
period $P\sim 0.5$ yr, but the model is worse than the best-fit parallax+orbit model by 
$\Delta\chi^2=118.3$, which is significant enough to exclude the xallarap interpretation.

\section{Characterizing the Lens}

\subsection{Physical Lens Parameters}

To uniquely determine the lens mass,
it is needed to estimate the angular Einstein radius in addition to the microlens parallax.
The angular Einstein radius is determined by
\begin{equation}
\thetae={\theta_*\over\rho}.
\end{equation}
We measure the normalized source radius $\rho$ from the analysis of the caustic-crossing 
parts of the light curve.  We note that the third peak was resolved with a sufficient 
coverage for the $\rho$ measurement.  See the upper panels of Figure~\ref{fig:four}.  
To determine $\thetae$, then, one needs to estimate the angular source radius $\theta_*$.

\begin{figure}
\includegraphics[width=\columnwidth]{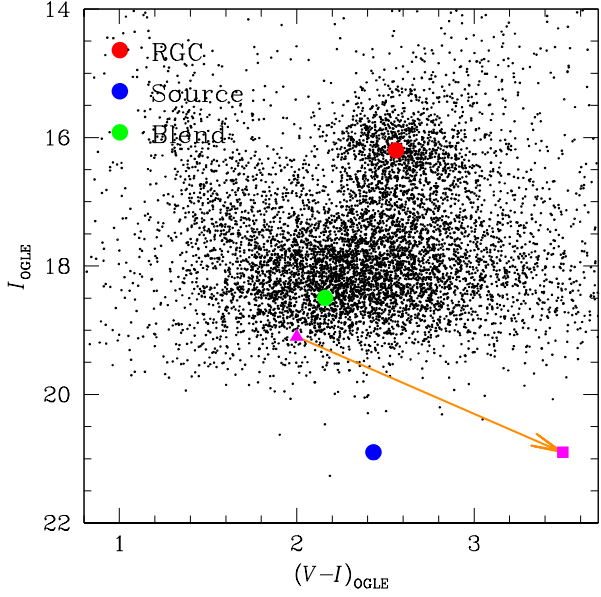}
\caption{
Source location with respect to the centroid of the red giant clump (RGC) 
in the color-magnitude diagram.
Also marked are the position of the blend.
The filled triangle and square dots denote the lens positions 
under the assumptions of no and full extinction, respectively. 
}
\label{fig:six}
\end{figure}

We determine the angular source radius based on the dereddened color $(V-I)_0$ 
and brightness $I_0$ of the source star.  For the color and brightness determinations, 
we use the method of \citet{Yoo2004}.  In this method, $(V-I)_0$ and $I_0$ are 
determined from 
the offsets in color $\Delta (V-I)$ and brightness $\Delta I$ with respect to the 
centroid of the red giant clump (RGC), for which the intrinsic color and brightness 
are known.  In Figure~\ref{fig:six}, we present the color-magnitude diagram of stars 
in the neighboring region around the source star.  The color-magnitude diagram is 
constructed based on the DoPHOT photometry of the $\mu$FUN CTIO data. It is aligned 
to the OGLE-III photometric system by shifting the clump magnitude according to the 
extinction, $A_I = 1.77$, and the reddening, $E(V - I) = 1.46$, toward the field 
based on the OGLE-III extinction map \citep{Nataf2013}.  We mark the 
positions of the RGC centroid and the source by a red and blue dots.  From the offsets 
in color $\Delta (V-I)=-0.39$ and magnitude $\Delta I=4.70$ and the known dereddened 
values of the RGC, $(V-I,I)_{{\rm RGC},0}=(1.06,14.41)$ \citep{Bensby2013, Nataf2013}, 
we find that the dereddeded color and brightness of the source star are $(V-I,I)_0=
(V-I,I)_{\rm RGC}+ [\Delta (V-I), \Delta I]=(0.93\pm 0.05,19.11\pm 0.01)$, indicating 
that the source is an early K-type main-sequence star. We then convert $(V-I)_0$ into 
$(V-K)_0$ using the $V-I$/$V-K$ relation of \citet{Bessell1988} and employ 
the color/surface brightness
relation of \citet{Kervella2004}
to find 
$\theta_* = 0.61 \pm 0.04\ \mu{\rm as}$.  
We estimate that the angular Einstein radius is 
\begin{equation} 
\thetae = 1.17 \pm 0.09\ {\rm mas}.  
\end{equation} 
Here we adopt the source distance  
that is estimated using the relation $D_{\rm S}=D_{\rm GC}/(\cos l+\sin l/\tan \phi)$ 
\citep{Nataf2013}, where $D_{\rm GC}=8160$ pc is the galactocentric distance, $l$ is 
the galactic longitude, and $\phi\sim 40^\circ$ is the angle between the semimajor 
axis of the bulge and the line of sight.  With $l=0.9^\circ$, the adopted source 
distance is $D_{\rm S}=8011$ pc.  In combination of the event timescale, the measured 
angular Einstein radius  yields the relative lens-source proper motion of 
\begin{equation}
\mu = {\thetae\over t_{\rm E}} = 
2.97 \pm 0.21\ {\rm mas}\ {\rm yr}^{-1}.
\end{equation}

With both measured $\pie$ and $\thetae$, the total mass $M=M_1+M_2$ is determined 
using the relation in Equation (2) and the masses of the individual components are 
determined by
\begin{equation}
M_{1}={M \over  1+q};\qquad M_{2}={qM \over 1+q}.
\end{equation}
The distance to the lens is determined by the relation
\begin{equation}
D_{\rm L}={{\rm au} \over \pie \thetae+\pi_{\rm S}},
\end{equation}
where $\pi_{\rm S}={\rm au}/D_{\rm S}$ represents the parallax of the source star. 
The projected separation between the lens components is determined by 
$a_\perp=sD_{\rm L}\thetae$.

In Table~\ref{table:four}, we list the physical parameters of the lens. 
We find that the lens is a binary composed of two M dwarfs with masses  
\begin{equation}
M_1=0.52 \pm 0.04\ M_\odot
\end{equation}
and 
\begin{equation}
M_2=0.42 \pm 0.03\ M_\odot.
\end{equation}
The estimated distance to the lens is 
\begin{equation}
D_{\rm L}=3.3 \pm 0.3\ {\rm  kpc}.
\end{equation}
The projected separation between the lens components is 
\begin{equation}
a_\perp = 6.4 \pm 0.5 \ {\rm au}.
\end{equation}
Also presented in Table~\ref{table:four} is the ratio of the transverse 
kinetic-to-potential energy ratio (KE/PE)$_\perp$.  The ratio is computed 
from the measured lensing parameters by 
\begin{equation}
\left( {{\rm KE}\over {\rm PE}}\right)_\perp = 
{(a_\perp/{\rm au})^3\over  8\pi^2(M/M_\odot)}
\left[ \left({1\over s} {ds\over dt}\right)^2+\left( {d\alpha\over dt}\right)^2 \right].
\end{equation}
The ratio should be less than unity to be a bound system, i.e.\ 
$({\rm KE/PE})_\perp \leq {\rm KE/PE}< 1.0$.
It is found that the determined value $({\rm KE/PE})_\perp=0.03$ meets this requirement.  
Due to the small lens-orbital effect, the ratio is small, probably due to the alignment 
of the lens components along the line of sight.

\begin{deluxetable}{lc}
\tablecaption{Physical lens parameters \label{table:four}}
\tabletypesize{\small}
\tablewidth{0pt}
\tablehead{
\multicolumn{1}{c}{Parameter} &
\multicolumn{1}{c}{Value}  
}
\startdata                                              
Primary mass          &  $0.52 \pm 0.04$ $M_\odot$  \\
Companion mass        &  $0.42 \pm 0.03$ $M_\odot$  \\
Projected separation  &  $6.4  \pm 0.5$ au          \\ 
Distance to the lens  &  $3.3  \pm 0.3$ kpc         \\  
(KE/PE)$_\perp$       &  0.03     
\enddata                                              
\end{deluxetable}

\subsection{Is the blend the lens?}

In Figure~\ref{fig:six}, we mark the location of the blend in the color-magnitude diagram.  
Then, a question is whether the blend is the lens itself.  The intrinsic 
color corresponding to the mass of the primary lens, $\sim 0.5 M_\odot$, is 
$(V-I)_{{\rm L1},0}\sim 1.9$.  Were it not for any extinction and reddening, the 
apparent color and  brightness of the primary lens would be 
$(V-I)_{\rm L1}\sim (V-I)_{{\rm L1},0}\sim 1.9$ and 
$I_{\rm L1}=M_{I,1} + 5 \log D_{\rm L} - 5 \sim 19.6$.  Here $M_{I,1}\sim 7.0$ 
represents the absolute magnitude corresponding to the mass of the primary lens.  
Considering that the primary is accompanied by a slightly less massive companion, 
the combined color and brightness of the lens would be  $(V-I,L)_{\rm L}\sim (2.0,19.1)$.  
If the blend is in the bulge, on the other hand, it would have experienced the same 
amount of extinction $A_I\sim1.8$ and reddening $E(V-I)\sim 1.5$ as those of the 
source star.  We mark the positions of the lens in the color-magnitude diagram under 
the assumptions of no and full extinction by a filled triangle and a square points, 
respectively.  Considering the distance to the lens of $D_{\rm L}\sim 3.3$ kpc, it 
is likely that the lens is inside the obscuring dust.  In this case, the lens on the 
color-magnitude diagram will be located on the line connecting the two points.  It 
is found that the blend is away from this line, suggesting that {\it the blend is 
not the lens}.  This line of reasoning is supported by the astrometric offset 
$\Delta\theta\sim 0.14"$ between the source position (measured from the difference 
image near the peak of the event) and the OGLE catalog position (which is dominated 
by the blend because it is $\sim 2$ magnitudes brighter than the source).

\begin{figure}
\includegraphics[width=\columnwidth]{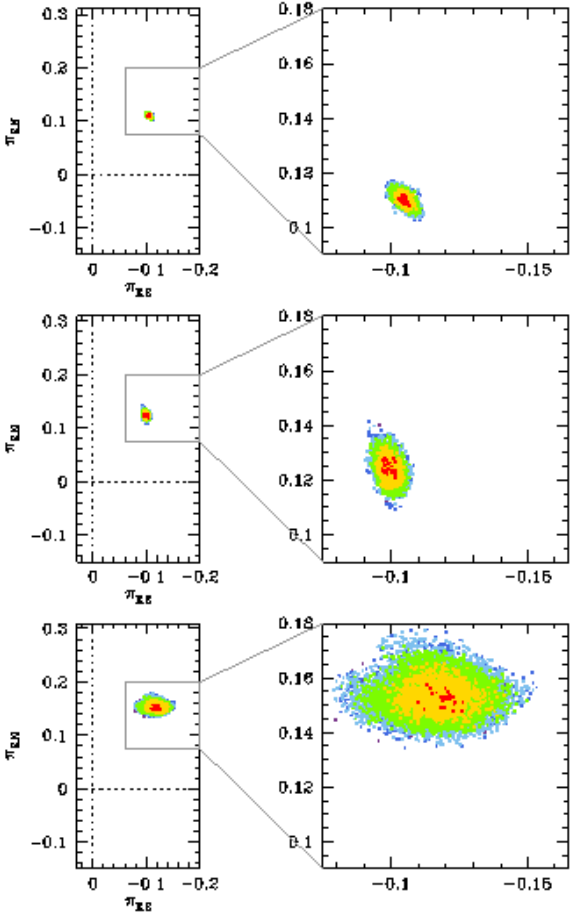}
\caption{
$\Delta\chi^2$ distributions of MCMC chains obtained from modeling runs based 
on different data sets.  The distribution in the upper panel is obtained based 
on all data. The distribution in the middle panel is based on the data where data 
points in the region $6890 < {\rm HJD}' < 6910$ are excluded. The distribution in 
the lower panel is based on the data with additionally excluded data points in the 
region $6850 < {\rm HJD}' < 6860$. Dots marked in different colors represent chains 
with $\Delta\chi^2 < 1$ (red), 4 (yellow), 9 (green), 16 cyan (magenta), and 25 
(blue). Right panels show the zoom of the regions enclosed by a box in the 
corresponding left panels.
}
\label{fig:seven}
\end{figure}

\section{Discussion}
 
The event OGLE-2014-BLG-0289 is very unusual in the sense that its light curve 
exhibits 5 peaks among which 2 were partially covered and the others are densely 
resolved.  In this section, we demonstrate that the quintuple peaks help to determine 
the microlens parallax with improved accuracy and precision.

For this demonstration, we conduct additional modeling runs with data sets 
where parts of the data points are excluded. In the first run, we exclude 
data points in the region $6890<{\rm HJD}'<6910$ to simulate the case where 
the fourth and fifth peaks were missed (`case 1'). In the second run, we 
additionally exclude data points in the region $6850 < {\rm HJD}' < 6860$ to 
simulate the case where the third peak was additionally missed (`case 2').

In Figure~\ref{fig:seven}, we present the $\Delta\chi^2$ distributions of MCMC 
chains in the $\pi_{{\rm E},E}$--$\pi_{{\rm E},N}$ parameter space obtained from 
the modeling runs with three different data sets.  Dots marked in different colors 
represent chains with $\Delta\chi^2 < 1$ (red), 4 (yellow), 9 (green), 16 cyan 
(magenta), and 25 (blue).  The upper panel is based on all data and the distributions 
in the middle and lower panels are for the `case 1' and `case 2', respectively. The 
right panels show the enlarged view of the regions enclosed by a box in the corresponding
left panels.

By comparing the distributions, we find that the microlens-parallax parameters 
determined based on the partial data sets differ from those based on the full data 
set by $\Delta (\pi_{{\rm E},N}, \pi_{{\rm E},E}) \sim(0.01, 0.01)$ for the `case 1' 
and $\sim (0.04, 0.02)$ for the `case 2'.  This indicates that the coverage of the 
peaks affects the accuracy of the $\pi_{\rm E}$ determination.  Furthermore, the 
uncertainties of the determined microlens-parallax parameters increase as fewer caustics 
are resolved, suggesting that the peak coverage also affects the precision of the 
$\pi_{\rm E}$ determination.  These results demonstrate that the resolution of the 
individual peaks provides important constraints on the determinations of the lens 
parameters.

We note that there was a discovery of an additional lensing event
with quintuple peaks. The event, Gaia16aye, showed a complex light curve
with 4 caustic crossings and a fifth brightening likely due to a cusp approach
\citep{Mroz2016, Wyrzykowski2017}.

\section{Conclusion}

We analyzed the binary-microlensing event OGLE-2014-BLG-0289. The light 
curve of the event exhibited very unusual five peaks where four peaks were 
produced by caustic crossings and the other peak was produced by a cusp 
approach. We found that the quintuple-peak features of the light curve 
enabled us to precisely and accurately measure the microlensing parallax 
$\pi_{\rm E}$. The three resolved caustics allowed us to precisely measure 
the angular Einstein radius $\thetae$. From the combination of $\pi_{\rm E}$ 
and $\thetae$, the physical parameters of the lens were uniquely determined. 
We found that the lens was a binary composed of two M dwarfs with masses 
$M_1 = 0.49 \pm 0.04\ M_\odot$ and $M_2=0.39 \pm 0.03\ M_\odot$ separated 
in projection by $a_\perp = 6.2 \pm 0.5$ au. The lens was located in the 
disk with a distance of $D_{\rm L} = 3.4 \pm 0.3$ kpc.

\acknowledgments
Work by C.~Han was supported by the grant (2017R1A4A1015178) of
National Research Foundation of Korea.
The OGLE project has received funding from the National Science Centre, Poland, grant 
MAESTRO 2014/14/A/ST9/00121 to A.~Udalski.  
OGLE Team thanks Profs.\ M.~Kubiak and G.~Pietrzy{\'n}ski for their
contribution to the OGLE photometric data set presented in this paper.
The MOA project is supported by JSPS KAKENHI Grant Number JSPS24253004, JSPS26247023,
JSPS23340064, JSPS15H00781, and JP16H06287.
Work by A.~Gould was supported by JPL grant 1500811.
Work by J. C. Yee was performed in part under contract with
the California Institute of Technology (Caltech)/Jet Propulsion
Laboratory (JPL) funded by NASA through the Sagan
Fellowship Program executed by the NASA Exoplanet Science
Institute.
We acknowledge the high-speed internet service (KREONET)
provided by Korea Institute of Science and Technology Information (KISTI).

\software{OGLE pipeline \citep{Wozniak2000, Udalski2003}, MOA pipeline \citep{Bond2001}, 
the DANDIA pipeline \citep{Bramich2008}, pySIS \citep{Albrow2009}, DoPHOT 
\citep{Schechter1993}.}

\end{document}